# Multiscale mechanical behavior of large arteries

**Authors :**
Claire Morin
   (1) Ecole Nationale Supérieure des Mines de Saint-Etienne, CIS-EMSE, SAINBIOSE, F-42023 Saint Etienne, France.
   (2) INSERM, U1059, F-42000 Saint Etienne, France.
   (3) Université de Lyon, SAINBIOSE, F-42000 Saint Etienne, France.

claire.morin@emse.fr

Witold Krasny
   (1) Ecole Nationale Supérieure des Mines de Saint-Etienne, CIS-EMSE, SAINBIOSE, F-42023 Saint Etienne, France.
   (2) INSERM, U1059, F-42000 Saint Etienne, France.
   (3) Université de Lyon, SAINBIOSE, F-42000 Saint Etienne, France.

Witold.krasny@emse.fr

Stéphane Avril
   (1) Ecole Nationale Supérieure des Mines de Saint-Etienne, CIS-EMSE, SAINBIOSE, F-42023 Saint Etienne, France.
   (2) INSERM, U1059, F-42000 Saint Etienne, France.
   (3) Université de Lyon, SAINBIOSE, F-42000 Saint Etienne, France.

avril@emse.fr

**Abstract:**
The mechanical integrity of arteries is of prime importance, for a proper oxygen and nutrients delivery to all organs. To optimize their mechanical properties, healthy arteries exhibit a complex hierarchical microstructure which ensures a sufficient compliance at low stresses and which stiffens at higher stresses, preventing over-dilatation. In this article, we propose a vast literature survey on how the mechanical properties of arteries are related to structural features. We first review the characteristics of arterial microstructure and composition and then we review the mechanical behavior of arteries. This allows evidencing, for the first time, the strong correlation existing between elastin and collagen contents within the arterial wall. However, bringing together most of the available mechanical tests on arterial walls shows the important variability of the mechanical arterial response. The presentation is organized at three different scales: at the tissue (macroscopic) scale, at the (micrometer) scale of the fiber networks, and at the (submicrometer) fibrillar and cellular scale.

**Glossary:**
*Autoclave:* pressure chamber used to carry out processes requiring elevated temperature and pressure different from ambient air pressure. In the sequel, autoclaving is used to separate soluble proteins (such as collagen) from non-soluble proteins (such as elastin).
*Constitutive relation*: in mechanics, material-specific relation between stress and strain that characterizes the response of the material to an applied mechanical loading.

*In vivo pre stretch and pre stress*: The *in vivo* state of the arterial tissue is not free of mechanical loading; in particular, it is subjected to an axial elongation and residual stresses.
*Staining*: auxiliary technique used in microscopy to highlight structures in materials. For biological tissues, it involves the use of structure-specific dyes.

**Nomenclature**: not needed

# I. Introduction

## *A. General information*

In the cardiovascular system, the main role of arteries is to ensure the circulation of blood from the heart to peripheral capillaries in order to deliver oxygen and nutrients to all tissues of the body. This physiological function can be maintained as far as the arterial system can receive spurts of blood from the left ventricle and distribute them as steady flow through peripheral capillaries. This implies that flow and pressure pulsations remain confined to the larger arteries which act as cushions through elastic dilatation and recoiling at every cardiac cycle (O'Rourke & Hashimoto 2007). In more details, during the systolic phase, the blood pressure rises and the large arteries distend and store blood; then, when the blood pressure falls during the diastolic phase, they recoil and discharge the previously stored blood to the smaller (muscular) arteries. This specific mechanical property of the large elastic arteries is usually referred to as the Windkessel effect (Wagner & Kapal 1952). Fulfilling these different functions involves ensuring the mechanical integrity of the arterial structure. Conversely, a failure in the mechanical integrity of arteries can have lethal consequences, due to a lack in oxygen supply to different organs. As a matter of fact, the World Health Organization reports that cardiovascular disorders are the first cause of death worldwide, killing 17.5 million of people yearly (i.e. 31% of deaths) (Sidloff et al. 2014; Santulli 2013). Many of these cardiovascular disorders are directly related to dysfunctions of the arterial mechanical role. For instance, an aneurysm is a permanent localized dilatation of an artery having at least a 50% increase in diameter compared to the expected normal diameter of the artery in question (Johnston et al. 1991). It is characterized by a modified and generally damaged arterial microstructure as well as by strongly modified mechanical load conditions applied on the arterial wall, endangering the mechanical integrity of the artery. When detecting an aneurysm, clinicians are confronted with the question of a surgical intervention, to replace the aneurysmal portion by a vascular prosthesis. Given that the mortality associated with this surgery oscillates between 3% and 5%, the quality of the criterion for surgery decision is of prime importance. Currently, surgery is clinically indicated by an invariable, purely geometrical criterion combining diameter and growth rate thresholds, representing a trade-off between surgery mortality and rupture (Davies et al. 2002; Fillinger 2007). However, this clinical decision between elective repair and conservative management of aneurysmal aortic tissue does not relate to any predictor of rupture risk (Gasser et al. 2010). As another example of a common cardiovascular pathology, atherosclerosis is a thickening and hardening of the arteries, characterized by the formation of a fatty, stiff plaque in the inner part of the arterial wall, which progressively obstructs the lumen, and therefore deteriorates the oxygen delivery to the organs. In some cases, bits of plaque may detach from the wall and may block the artery downstream, eventually causing a heart attack or stroke.
Being able to assess the mechanical integrity of a blood vessel in relation to the prevailing loading conditions is therefore essential for adapting and improving the management of these cardiovascular disorders (deBotton & Oren 2013).

## *B. Structure-property relations*

The mechanical integrity of biological structures primarily depends on the mechanical loading to which the structure is subjected, the specific geometry of the structure, and the mechanical properties of the tissue making up the structure. At the tissue scale, the mechanical properties of arteries appear as extremely scattered and were shown to vary during aging or due to pathologies, but also across the considered organs or species (Lally et al. 2004; Weizsäcker et al. 1983; Holzapfel et al. 2005). However, the aforementioned variability of mechanical properties may mainly be governed by the variability of the associated microstructural and compositional properties of the tested tissues (Holzapfel et al. 2005).

Due to its biological nature, the arterial composition and microstructure are not invariant over time, but actually change due to biological processes called growth and remodeling (Gibbons & Dzau 1994; Humphrey 1995). The latter depend not only on biochemical conditions but also on the mechanical loading itself (Leung et al. 1976; Humphrey 2006; Humphrey et al. 2014), such that the tissue actively adapts to its bio-chemo-mechanical environment. Tissue adaptation is thought to be governed by the different populations of cells in the arteries which try to reach a homeostatic mechanical state (Humphrey 2008). In the occurrence of cardiovascular pathologies or upon aging, abnormalities in the remodeling process may severely harm the mechanical integrity of the system (Jacob et al. 2001). For instance, non-homeostatic blood pressure conditions (as is the case in hypertension) affect the structure of the arterial wall, which is found different as compared to arteries subjected to homeostatic conditions (Berry & Greenwald 1976; Jacob et al. 2001). These altered conditions may lead to accelerated fatigue and failure of the different arterial wall constituents, such as the elastic lamellae; and therefore to a stiffening of the arterial wall and to transmission of the pulsatile nature of the blood flow to smaller vessels and organs. Also, in aortic aneurysms for instance, (mechanical) damage to or (chemical) degradation of the elastic fibers, in combination with the loss of smooth muscle function is a common early contributor to the formation or expansion of all aneurysms. Then, both remodeling of collagen (i.e. turnover) and inflammation (i.e. invasion of mononuclear cells such as lymphocytes T and B) can play fundamental roles in dictating rates of enlargement as well as the (increased) probability of aneurysm rupture (Jacob et al. 2001). All these examples show that the vascular pathologies may be translated at a given time instant into microstructural changes, whether composition, arrangement, or mechanical properties of the individual constituents. Still, up to now, the mechanical and microstructural investigations were carried out separately and the bridges between the two communities remained scarce. However, improving diagnosis and treatment of cardiovascular pathologies necessarily implies to account for the structure-property relations (Humphrey 2009).

In this article, we propose to review relevant information about microstructure and mechanics of large elastic arteries, which could be regarded as useful for subsequent definition of structure-relation properties. The present review seeks for general patterns governing the multiscale microstructure organization, characterization, and composition (see section 2), as well as patterns governing its multiscale mechanical properties (see section 3), valid over the widest range of species and organs possible. This review is limited to the healthy large elastic arteries.

## II. Characterization of the arterial tissue's structure

Arteries are characterized by a great diversity and variability. Depending on the species or on their location in the organism, they have to sustain and regulate different levels of blood pressure. Still, the hierarchical structure of the arterial wall remains unchanged among the vertebrate kingdom. The arterial wall is made up by three main elementary constituents, namely elastin (see Figure 1(j)), collagen (see Figure 1(h)), and water, as well as by different other organic molecules (e.g. glycosaminoglycans, proteoglycans). These elementary constituents are arranged in morphological hierarchical structures occurring in most of the large arteries of vertebrates, and described by means of the following three levels:

- At a characteristic length of several millimeters, the macrostructure of the arterial wall is characterized by varying thicknesses and mechanical properties depending on the precise biological function of the considered arterial wall (see Figure 1(a)). This arterial macrostructure is composed, at a few hundreds of micrometers scale, of three concentric tunicae: the intima, media, and adventitia, see Figure 1(b). Each of these layers has a specific morphology as well as a

- specific function within the arterial wall. This scale will be referred to as the macrostructure of the arterial wall.
- At a characteristic length of some micrometers to some tens of micrometers, each layer is made up by an arrangement of cells embedded in different imbricated fibrous elastic and collagenous networks (see Figure 1(c-f)). This will be referred to as the arterial microstructure.
- Zooming at these fibrous networks reveals the ultrastructure of the fibers, made of an arrangement of crosslinked fibrils, at a hundred of nanometers scale, see Figure 1(g, i).

Understanding the arrangement of the different constituents within the arterial wall has been the source of an impressive amount of scientific publications; and the emergence in the last fifteen years of multiphoton microscopy has led to a better understanding of the constituent arrangement and of the relation to their mechanical function. After a brief overview of the diverse observation techniques, the multiscale description of the arterial wall microstructure will be reviewed in details. As a second step, the qualitative hierarchical description of arteries is complemented by the extraction of quantitative parameters characterizing this microstructure, such as arterial composition, fibers orientation and shape, etc.

## A. *Multiscale observation techniques of the arterial structure*

The hierarchical character of arteries has been revealed by the use of different microscopy techniques, offering a wide range of resolutions, as well as different imaging characteristics (in-depth resolution, need for prior staining, etc.).

### 1. Macrostructure (100μm-5mm)

The composite structure of the artery has been originally revealed by histology studies using optical microscopy. The latter imaging technique is dedicated to the study of the microscopic anatomy of cells and tissues. It uses visible light and a system of lenses to magnify images of biological samples. The specimens are previously sectioned (cut into a thin cross section with a microtome), stained, and mounted on a microscope slide.

### 2. Microstructure (5-100 μm)

Different microscopy techniques may be used to visualize the arrangement of the vascular microstructure at the micrometer scale. As already mentioned, multiphoton microscopy (van Zandvoort et al. 2004) is a fluorescence imaging technique that uses near-infrared excitation light and that enables in-depth imaging of living tissue with an imaging maximal depth of the order of one millimeter, therefore needless of prior tissue sectioning. The multiphoton absorption enables suppression of the background signal. When imaging biological tissues, the signal of collagen is generated from the second harmonic generation, while the elastin signal is recovered by auto fluorescence at the right excitation wavelength, before being collected by two bandpass filters. Multiphoton microscopy can also reveal vascular cells such as fibroblasts or smooth muscle cells (O'Connell et al. 2008) upon prior staining for fluorescence. Multiphoton microscopy can be upgraded by the addition of a spatial pinhole placed at the confocal plane of the lens. This technique is then called confocal microscopy or confocal laser scanning microscopy (Voytik-Harbin et al. 2003). In such a configuration, the optical resolution and contrast are increased while out-of-focus light is eliminated, enabling the reconstruction of three-dimensional structures from the obtained images by collecting sets of images at different depths. As the light from sample fluorescence is blocked at the pinhole, the benefit of increased resolution decreases the signal intensity; this induces long exposures of the biological samples to the imaging beam.

Diffusion tensor imaging (Vilanova et al. 2006) relies on a different technology: it uses the diffusion of water molecules to generate contrast in magnetic resonance images. As the molecular diffusion of water in tissues is constrained, it can reflect interactions with many obstacles when tracked, such as macromolecules, fibers, and membranes. This tracking provides a mapping of diffusion patterns therefore revealing microscopic details about the tissue architecture.

Tomographic imaging of biological tissues provides section views of organic components through the use of various penetrating waves (e.g. X-ray). Tomographic slices can be reconstructed into 3D views by means of specific reconstruction techniques (Fujimoto et al. 1999).

Finally, the vascular microstructure can also be imaged using polarized light microscopy (Gasser et al. 2012). Polarized light is obtained by means of a polarizer oriented at 90° for blockage of directly transmitted light. When imaging biological tissues under polarized light, previously stained collagen expresses different interference colors influenced by the fiber thickness but also by the packing of the collagen fibers.

### 3. Ultrastructure

At the scale of a few tens to hundreds of nanometers, the resolved structure of the different fiber networks can be revealed by means of scanning electron microscopy. This technique uses a beam of accelerated electrons as a source of illumination. The higher resolving power of scanning electron microscopes originates in the wavelength of electrons being up to 100,000 times shorter than that of visible light photons used in optical microscopy. Prior acid and elastase digestion of the tissue can be realized in order to expose particular cells.

## B. Universal hierarchical structure of the arterial tissue

Use of these different microscopy techniques allowed revealing the hierarchical structure of arteries.

### 1. Macrostructure

At the macroscopic scale, the artery is a composite cylindrical structure made of three concentric layers: the adventitia, the media, and the intima, as seen on Figure 1(b), an electron micrograph obtained by (Ratz 2016). Each of these layers is characterized by specific microstructures, specific thicknesses (Wolinsky & Glagov 1967), different mechanical properties (Holzapfel et al. 2005), and different structural and biological functions (O'Connell et al. 2008). The relative proportions of layer thicknesses differ in particular between proximal and distal regions (Canham et al. 1989).

First, the most inner layer of the arterial wall is the tunica intima, made of the endothelium and of an internal elastic lamina. The endothelium is a monolayer of endothelial cells, lining the luminal surface of blood vessels, as seen on Figure 1(c), obtained from (Ives et al. 1986) by means of an inverted phase optical microscope. It plays the role of an interface between blood vessel walls and the blood flow (Ives et al. 1986) being therefore subjected to both fluid shear stress and pressure-induced strain components of the flow (Ives et al. 1986). In particular, the effect of shear stress on endothelial cell morphology and functions has been exclusively studied, highlighting that the endothelium shows a tendency toward parallel alignment with the principal axis of strain; and that endothelial cell morphology is closely related to its cytoskeletal structure (Ohashi & Sato 2005). Moreover, the mechanical forces coming from blood flow have been proposed as causative factors in cardiovascular diseases, such as the atherosclerosis, and have been implicated in modulating endothelial cell morphology and function (Ives et al. 1986). The internal elastic lamina (not shown on Figure 1) follows the endothelium, and is known to provide structural cohesion and support for axial pre-tension (Farand et al. 2007; Timmins et al. 2010).

Then, the tunica media, the second concentric layer making up the arterial wall, is a concentric set of superimposed medial lamellar units, as reconstructed on Figure 1(f1) from a stack of confocal microscope images. A single medial lamellar unit, as shown on Figure 1(d), obtained from (O'Connell et al. 2008) by means of scanning electron microscopy, consists in a row of overlapping smooth muscle cells embedded on the upper and lower sides by two concentric elastic lamellae. The overlapping muscle cells lie parallel to tangential planes of the orthoradial direction. The number of lamellar units in the media of adult mammalian aortas has been shown to be nearly proportional to the aortic radius regardless of species or of variations in measured wall thickness (H. Wolinsky & Glagov 1967). The tunica media is therefore, from a morphological point of view, a composite material organized periodically: the radial transmural disposition of cells and matrix fibers on transverse sections of the media in well-developed aortas is proved to be cells, elastic lamellae, cells, etc.

Finally, the tunica adventitia, the third and most outer concentric layer of the arterial wall, is made of an arrangement of collagen bundles and few elastic fibers, as seen on Figure 1(e1) obtained from reconstruction of a stack of confocal microscope images, together with embedded fibroblasts.

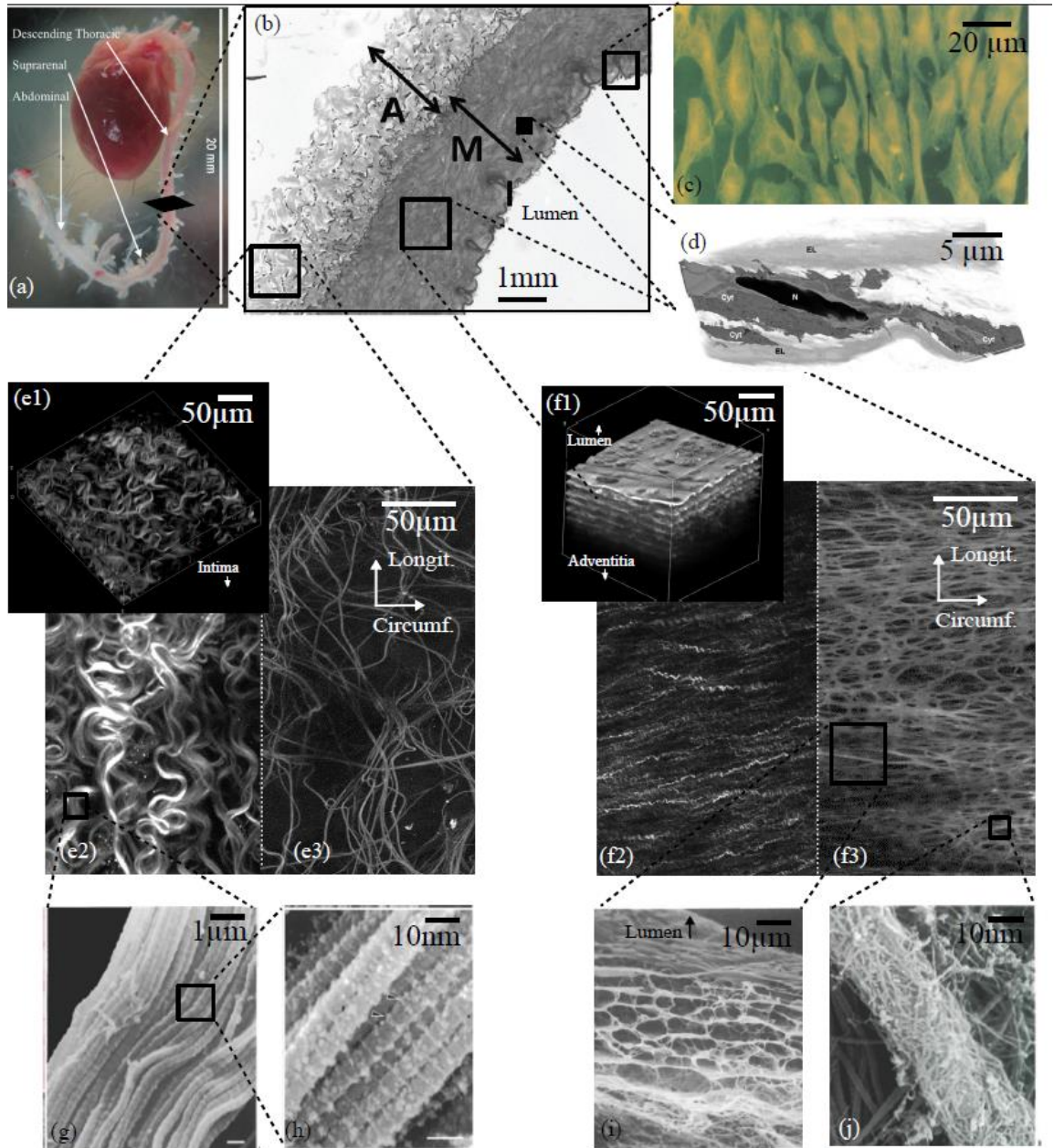

Figure 1: Hierarchical structure of large elastic arteries: (a) macroscopic view of large arteries, taken from (Keyes et al. 2011); (b) at the millimeter scale, the arterial wall is made of three concentric layers, the adventitia tunica (A), the tunica media (M) and the intima tunica (I); electron micrograph from (Ratz 2016); (c) at the micrometer scale, the intima is made of a continuous layer of endothelial cells (inverted phase microscope image from (Ives et al. 1986)); the media is made of an arrangement of medial lamellar units, as seen on (d) by means of scanning electron microscopy (taken from (O'Connell et al. 2008)), and (f1) from confocal microscopy stack of images; zooming further on the lamellar unit, confocal microscopy allows to distinguish (f2) the collagen fibers and (f3) the elastin network; the adventitia is made of (e1) an arrangement of collagen bundles (also e2) and elastin fibers (also e3); finally, the collagen fibers are made of a (g) staggered arrangement of collagen fibrils (scanning electron microscope image taken from (Ushiki 2002)), themselves made of an (h)

arrangement of crosslinked collagen molecules (scanning electron microscope image taken from (Ushiki 2002)); while the elastin network is made of (i) elastic fibers, lamellae, and struts, themselves made of an (j) arrangement of elastin and crosslinking molecules (scanning electron microscope image taken from (Ushiki 2002)). Images (e1)-(e3), (f1)-(f3) were obtained by a confocal bi-photon microscope (IVTV Platform, ANR-10-EQPX-06-01, FR) imaging a rabbit carotid artery.

## 2. Microstructure

At the micrometer scale, microscopy techniques reveal the precise morphology of the different fiber networks that exist in each arterial layer. Starting from the arterial lumen, in the endothelium, the actin filaments (F-actin) are one of the major cytoskeletal structures of the endothelial cells (Ookawa et al. 1992); this actin network also exists in the smooth muscle cells. They are organized in bundles and are usually grouped in the central part of the cells. Noticeably, the redistribution of F-actin filaments within the cells is one of the early cellular responses to the onset of shear stress: when experiencing low-shear forces coming from the blood flow, F-actin filaments localize at the periphery of the endothelial cells (Ookawa et al. 1992); whereas when experiencing high shear forces, F-actin bundles are observed in the central part of the elongated cells. In parallel, the cells orient in the direction of the applied flow. A continuous sheet of elastin, called the internal elastic lamina (Farand et al. 2007; Timmins et al. 2010), surrounds the endothelium and marks a frontier between the endothelium and the tunica media. It takes the form of a dense elastin sheet equipped with a longitudinal network of elastic fibers coated on its continuous surface.

Then, as mentioned earlier, the medial lamellar unit is made of an arrangement of elastic fibers, vascular smooth muscle cells, and collagen fibers, which are now successively described. From a morphological point of view at the micrometer scale, medial elastin takes three different forms:

- Firstly, lamellae, composed of a dense meshwork of elastic fibers oriented circumferentially (Farand et al. 2007; Timmins et al. 2010). They form a periodical concentric separation between medial lamellar units containing the vascular smooth muscles cells, see Figure 1(f1) and Figure 2(E1). These lamellae show large, round, reinforced fenestrations (see Figure 2(E2), also visible on the upper lamella of Figure 1(f1)), and house the superior and inferior anchorages of smooth muscle cells, allowing them to weave through the tunica media (Dingemans et al. 2000);
- Secondly, thick radial elastin struts provide structural cohesion to the overall medial lamellar unit, while preserving an angular 20° tilt of smooth muscle cell orientation, see Figure 2(E3) (Dingemans et al. 2000; Koch et al. 2014; Tsamis, Phillippi, et al. 2013);
- Finally, thin elastic radial fibers take the form of ridges (Dingemans et al. 2000) or protruding ribs (Raspanti et al. 2006), connecting the smooth muscle cells to both lamellae, see Figure 2(E4).

Through their actively adaptive plasticity, the vascular smooth muscle cells are responsible for the regulation and maintenance of blood flow and for the regulation of stress across the arterial wall thickness. For this reason, vascular smooth muscle cells house actin-myosin filaments (myofilament) that permit rapid stress-development, sustain stress-maintenance and vessel constriction (Ratz 2004). The relaxed smooth muscle cell is a very long and thin (fusiform) structure (Ratz 2004) with a high surface area, and an ellipsoidal shape of its nuclei (O'Connell et al. 2008). Upon contraction, the vascular muscle cells can undergo dramatic shortening accompanied by shape change, surface rearrangements, and a loss of volume that is regained upon relengthening. As for their spatial positioning among the constituents of the medial lamellar unit, it shows noticeable characteristics, namely the cells weaving throughout the interlamellar elastin framework, resulting in approximately 20° radial tilt (O'Connell et al. 2008).

Collagen is also organized in different morphologies within the tunica media. Firstly, interlaced bundles of type IV collagen microfibrils (oxytalan fibers) of the immediate pericellular matrix contribute, along with elastin radial struts and interlamellar elastin protrusions, to smooth muscle cell cohesion and fixation on the lamellae, see Figure 2(C2) (Clark & Glagov 1985; Dingemans et al. 2000). It is understood, that the smooth muscle cells preferentially adhere to these ill-defined streaks rather than directly to the solid lamellae, see Figure 2(SMC) (Dingemans et al. 2000); secondly, wavy collagen fiber bundles are interposed between the facing elastin systems within the fibrous regions between cell layers, see Figure 2(C1) and Figure 1(f2) (Clark & Glagov 1985). These collagen fibers are oriented circumferentially (O'Connell et al. 2008; Timmins et al. 2010; Roy et al. 2010; Hill et al. 2012) and are closely associated with the elastic lamellae (Dingemans et al. 2000) but not with the smooth muscle cells. Upon pressure, these medial collagen fibers decrimp and stretch to prevent over distension of the vessel. Collagen takes also the form of membranes enveloping the smooth muscle cells; see Figure 2(C3).

Finally, the most outer arterial layer, the tunica adventitia, is made of an arrangement of networks of elastin and collagen with embedded fibroblasts. In the adventitia, elastin takes the form of a low density meshwork made of variously oriented fibers showing bifurcations (transversely oriented segments), with a dominant longitudinal direction (Chen et al. 2011; Chen et al. 2013), see Figure 1(e3). Contrarily, collagen fibers pack into thick bundles of 10 to 30 fibers, folded (crimped) under the *in vivo* pre-stress and pre-stretch conditions. These wavy bundles are oriented helicoidally about the 45° angle and show a negligible transmurality (Roy et al. 2010; Rezakhaniha et al. 2012; Schrauwen et al. 2012), see Figure 1(e1) and (e3). For an excised cut open tissue, the crimping appears considerably more important than in the excised cylindrical configuration, and the bundles are oriented closer to the axial direction (Tsamis et al. 2013; Phillippi et al. 2014; Koch et al. 2014; D'Amore et al. 2010). The adventitial collagen network is capable of undergoing important morphology rearrangements under mechanical load; namely decrimping, stretching, and reorientation in the load direction, with amplitudes that can exceed affine predicted reorientation (Billiar & Sacks 1997; Chandran & Barocas 2006). These changes in morphology are known to be linked to the important non linearity of the mechanical response, and give rise to a very important material stiffening under large strain (Chen et al. 2011; Schrauwen et al. 2012). These aspects are further detailed in the mechanics section 3 of this article. The adventitial microstructure is also composed of fibroblasts (not shown on Figure 1), which are mechanosensitive cells, active in the arterial remodeling process and arranged circumferentially about collagen bundles (Esterly et al. 1968).

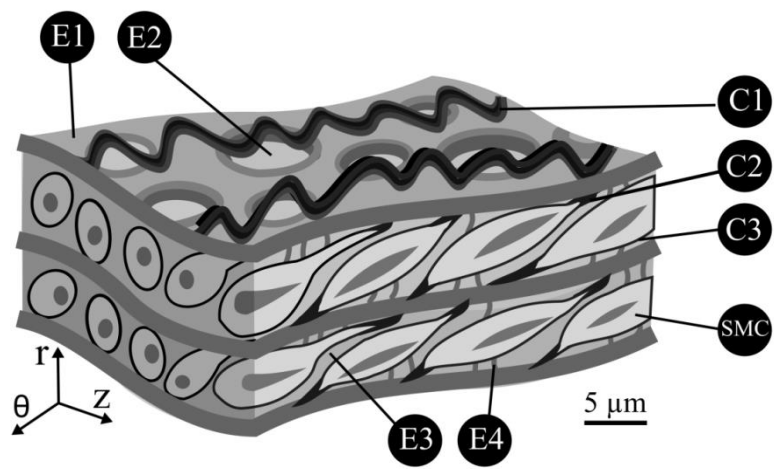

Figure 2: Schematic representation of the medial lamellar unit morphology. (E1) elastin lamella, (E2) fenestration of the elastin lamella, (E3) elastin interlamellar struts, (E4) elastin surface ridges, (SMC) smooth muscle cells, (C1) collagen fibers, (C2) cohesive collagen microfibril bundles, (C3) collagen envelope of SMCs. Inspired from illustrations by (Dingemans et al. 2000; O'Connell et al. 2008)

### 3. Ultrastructure

At the hundreds of nanometers scale, the structure of elastic and collagen fibers is revealed, as well as the existence of crosslinks between and within these fibers. The morphology of the elastic lamellae presents a fibrous texture suggestive of a "criss–crossed", delicate filamentous scaffold (Ushiki 2002; Raspanti et al. 2006). The elastic fiber and elastin meshwork however are made of 0.2 µm thick elastin fibrils and microfibrils that run in various directions, see Figure 1(i) and (j). Those microfibrils are coated together within an elastic fiber by proteoglycans and glycosaminoglycans

(Ushiki 2002). At the same scale, collagen bundles are made of closely packed, parallel, thin collagen fibrils, with a characteristic diameter of 30 to 100 nm, with altering number of coated fibrils depending on the region in the bundle (Ushiki 2002; Raspanti et al. 2006). In the adventitia, these bundles are thicker than in the media due to a higher number of constituting fibrils, see Figure 1(g). The fibrils present a regular, orthogonal lattice of surface-bound proteoglycans (Raspanti et al. 2006; Berillis 2013), see Figure 1(h). Finally, at the nanometer scale, the cohesion between fibrils constituting elastic or collagen fibers is realized by crosslinking proteoglycans, formed by covalent bonding between acid glycosaminoglycans and proteins (Eisenstein et al. 1975). At the microfibrillar level, collagen and elastin are cross-linked by a unique mechanism based on aldehyde formation from lysine or hydroxylysine side chains (Eyre et al. 1984; Sáez et al. 2016).

## C.   *Quantitative characterization techniques of the arterial microstructure*

Many tests were performed over the last 70 years to determine the relative amount of elastin, collagen, and water, among different arteries, species, and at different ages, resulting in a very broad variety of arterial compositions. The relative amounts of collagen and elastin can be determined at different scales, depending on the chosen technique: chemical methods provide average weight fractions of collagen and elastin over a millimeter-sized dehydrated defatted arterial sample, while histochemical methods allow determining the volume fractions of collagen and elastin fibers in the arterial tissue.

### 1.   Collagen and elastin weight fractions at the macroscopic scale

Corresponding experimental endeavors involve primarily the following steps:
- The arterial samples are generally excised from freshly sacrificed animals and analyzed directly after excision, or wrapped airtight in plastic films and stored at -20°C (Brüel & Oxlund 1996; Looker & Berry 1972). In some cases (Leung et al. 1977; Berry & Greenwald 1976; Myers & Lang 1946), the adventitia is removed by careful dissection, and only the compositions of media and intima are determined.
- Defatting the arterial wall is performed by means of successive immersions in acetone and ether, for different durations depending on the exact protocol (Fischer & Llaurado 1966; Harkness et al. 1957; Grant 1967; Neuman & Logan 1950; Feldman & Glagov 1971);
- The dehydration procedure consists in either drying to constant weight in vacuum (Harkness et al. 1957; Looker & Berry 1972; Farrar et al. 1965; Hosoda et al. 1984; Berry & Greenwald 1976; Feldman & Glagov 1971; Spina et al. 1983), or drying to constant weight in an oven for a few hours at a temperature higher than 50 °C (Grant 1967; Fischer & Llaurado 1966; Lowry et al. 1941; Neumann & Logan 1950; Leung et al. 1977). As a result, water represents about 70%-80% of the wet weight of the aortic tissue (Looker & Berry 1972; Fischer & Llaurado 1966; Dahl et al. 2007).
- Although different methods exist for the detection and estimation of collagen weight fraction in a biological sample, the determination of hydroxyproline in the tissue seems to be the most widely used method. Hydroxyproline is an amino acid, whose content varies between 13.1% in collagen type I to 17.4% in collagen type III (Etherington & Sims 1981). As elastin also contains hydroxyproline, collagen needs first to be removed from the tissue by autoclaving. This separation method between elastin and collagen accounts for the non-soluble property of elastin. More precisely, a small piece of the dry defatted tissue is autoclaved twice for 3 hours at 1 bar[1], then washed with water and centrifuged; the resulting extracts are dried out (Neuman & Logan 1950). Then, the following steps (with possible slight adaptations) are followed for hydroxyproline determination: (i) hydrolysis at 3.4 bar for 3 hours with 6 mol.L$^{-1}$ hydrochloric acid to release the hydroxyproline from peptide linkage, (ii) oxidation with sodium peroxide, and (iii) formation of a reddish purple complex with p-dimethylaminobenzaldehyde (Harkness et al. 1957; Looker & Berry 1972; Farrar et al. 1965; Berry & Greenwald 1976; Neumann & Logan 1950). Chloramine T together with Ehrlich's reagent for colorimetric determination were preferred for use as an oxidant instead of sodium peroxide (Stegemann & Stalder 1967; Hosoda et al. 1984; Brüel & Oxlund 1996; Han et al. 2009). The collagen content follows from a multiplicative correction of the hydroxyproline content; this correction factor varies among the sources between 7.14, 7.46, or 7.8 depending on the exact content of hydroxyproline in the different collagen types (Neuman & Logan 1950; Feldman & Glagov 1971; Fischer & Llaurado 1966; Grant 1964; Grant 1967; Harkness et al. 1957; Berry & Greenwald 1976).

---

[1] Pressure and duration of autoclaving were varied by different experimentalists: (Grant 1964; Grant 1967) opted for 6h at 1 bar; (Feldman & Glagov 1971) for 6h at 2.75 bar; (Fischer & Llaurado 1966) for 18h at 1-1.4 bar followed by a second for 3h; (Harkness et al. 1957; Looker & Berry 1972; Farrar et al. 1965; Berry & Greenwald 1976) for 6h at 2 bars; (Lowry et al. 1941) for 4h at 3.5 bars.

- After autoclaving the arterial sample, the insoluble elastin remains in the residue, and its content is determined by a gravimetric method (Looker & Berry 1972; Berry & Greenwald 1976; Farrar et al. 1965; Leung et al. 1977; Hosoda et al. 1984; Lowry et al. 1941; Myers & Lang 1946; Feldman & Glagov 1971). In short, this method consists in purifying the residue remaining after collagen extraction, first by heating for 45 min at 100°C with a 0.1 mol.L$^{-1}$ sodium hydroxide solution (Lansing et al. 1952), then by washing with water, dehydrating with acetone, and finally drying to constant weight in a hot air oven (Grant 1967) or in vacuum over phosphorous pentoxide (Harkness et al. 1957; Looker & Berry 1972) and weighing. Other methods for elastin determination include the same procedure of hydroxyproline determination as for collagen (Fischer & Llaurado 1966; Harkness et al. 1957; Grant 1967; Grant 1964; Leung et al. 1977), and the determination of elastin-specific crosslinking amino acids (desmosine and isodesmosine) (Han et al. 2009).

Finally, the volume fractions of cells and/or the cell content were also determined in several studies: dosage of DNA gives access to the total cell content of arterial tissue, being of 200 million of cells per milligram of dry defatted tissue in adult rats (Dahl et al. 2007), while histological observations with prior cell staining provide access to the volume fraction of some specific cell populations (O'Connell et al. 2008; Tonar et al. 2008).

## 2. Volume fraction of fibers at the macroscopic scale

Histochemistry consists in studying the chemical constituents of a tissue by means of staining reactants. The arterial tissue is first fixed with paraffin and then stained with different dyes. In particular, collagen stains include (blue or green) Masson's trichrome (Phillippi et al. 2014; Anidjar et al. 1990; Dahl et al. 2007), Picrosirius red (Phillippi et al. 2014), (yellow) Safranin-O staining (Azeoglu et al. 2008), (pink) eosin (Carmo et al. 2002), Van Gieson staining, which is a mixture of picric acid and acid fuchsin, staining collagen in red, nuclei in black, and cytoplasm in yellow (Carmo et al. 2002; Cattell et al. 1996). Elastin stains include (blue-black) Verhoeff's staining (Tonar et al. 2003; Phillippi et al. 2014; Dahl et al. 2007), permanganate-bisulfite-toluidine blue reaction (Clark & Glagov 1985; Fischer 1979), (red) orcein staining (Hosoda & Minoshima 1965; Anidjar et al. 1990; Scarselli & Repetto 1959; Scarselli 1959), Weigert's (blue-black) resorcin-fuchsin staining (Hayashi et al. 1974). Histochemical methods also allow cells staining: toluidine blue stains nucleic acids in blue, hematoxylin (also called Weigert's iron hematoxylin stain) stain cells nuclei in dark blue, eosin stains in pink the cell's cytoplasm (Carmo et al. 2002); finally, Movat's stain allows to identify glycosaminoglycans in blue (Dahl et al. 2007).

The stained histological section is observed under a microscope and image processing techniques provide access to the searched parameters. E.g., several methods have been proposed to analyze the statistical fiber orientation based on microstructure imaging; they involved Hough transforms (Chaudhuri et al. 1993; Karlon et al. 1998), structure tensor-based texture analysis (Rezakhaniha et al. 2012), direct fiber tracking (Pourdeyhimi 1999; Mori & van Zijl 2002; Rezakhaniha et al. 2012; Hill et al. 2012; Ghazanfari et al. 2012), or 2D Fast Fourier transform (Ayres et al. 2006; Ayres et al. 2008; Timmins et al. 2010; Schriefl et al. 2012; Polzer et al. 2013).

## D. Quantitative assessment of the arterial microstructure

### 1. Universal pattern for the arterial composition

The mechanical properties of arterial tissues depend primarily on the tissue composition and on the arrangement of the constituents of its microstructure. Since access to the arterial composition is complex *in vivo*, we here propose to seek for such a general composition rule, valid over a wide variety of organs, species and ages. To our best knowledge, no such relation governing the arterial composition has ever been proposed. At most, correlations between the arterial composition and some other characteristics of the tissue have been established, but generally limited to very restricted sets of data. We here collected the weight fractions of elastin and collagen in various large elastic arteries from a great variety of species and ages: from rats to human, from the abdominal to the carotid arteries, etc. However, since important modifications in the arterial composition occur in the perinatal and early childhood periods (Bendeck & Langille 1991), as well as in aging organisms (Tsamis, Krawiec, et al. 2013; Schlatmann & Becker 1977)[2], we restricted our selection to non-aging adult organs. We plot the collagen weight fraction as a function of the elastin weight fraction, for cases where those contents are related to the media and intima only (see Figure 3) or to the whole arterial thickness (see Figure 4). It is interesting to notice that collagen and elastin contents strongly correlate mutually, whether only the intima-media is considered or the entire thickness of the arteries. A closer look into the resulting graphs shows that, as a rule, the thoracic aortas are characterized by a larger content in elastin as compared to the abdominal aortas, which contain more collagen (compare the filled markers in Figures 3 and 4 to the empty ones). This is in agreement with observations made on more restricted arterial tissue origins and numbers by (Harkness et al. 1957; Sokolis 2007; Halloran et al. 1995; Cheuk & Cheng 2005): the elastin content drops when passing from proximal to distal aortas. In parallel, it was also observed that the number of lamellar units also drops in more distal aortas (Wolinsky & Glagov 1969; D. P. Sokolis et al. 2002). These changes in the arterial composition affect the mechanical behavior, as reviewed in section 3.

Such a good correlation between elastin and collagen content could be used to determine the composition of healthy adult arteries, or as a check of the healthy compositional state of an arterial segment.

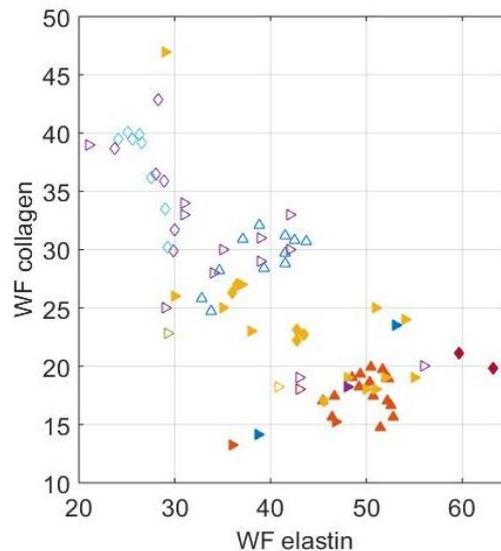

**Figure 3: Composition of the media and intima layers of different arteries stemming from rabbit (upward pointing triangles), rats (diamonds), and human (right pointing triangles). Collagen and elastin weight fractions (WF) are related to the dry defatted weight of the arterial tissue. Filled markers correspond to thoracic arterial segments, while empty**

---

[2] In the perinatal period, a rapid accumulation of both elastin and collagen has been observed followed by a marked postnatal increase in arterial pressure, see (Bendeck & Langille 1991). Upon aging, the absolute collagen content is increased, while the absolute elastin content remains constant, although the elastin becomes fragmented (Tsamis, Krawiec, et al. 2013; Schlatmann & Becker 1977).

markers correspond to abdominal arterial segments as well as one pulmonary trunk tissue and one arch tissue. Data were taken from (Leung et al. 1977; Berry & Greenwald 1976; Brüel & Oxlund 1991; Han et al. 2009; Andreotti et al. 1985; Feldman & Glagov 1971; Apter et al. 1966; Tonar et al. 2003; Spina et al. 1983)

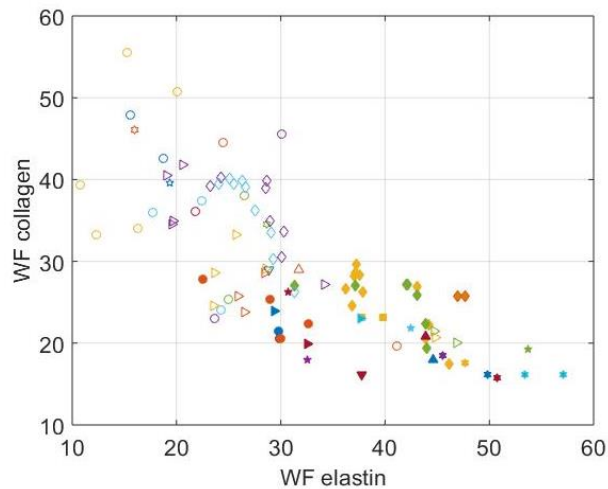

Figure 4: Composition of the whole arterial segment of different arteries stemming from pigs (hexagrams), dogs (circles), sheep (downward pointing triangles), rats (diamonds), goats (upward pointing triangles), bovines (square), puppies (pentagrams), and humans (right pointing triangles). Collagen and elastin weight fractions (WF) are related to the dry defatted weight of the arterial tissue. Filled markers correspond to thoracic and arch arterial segments, while empty markers correspond to abdominal and other arterial segments. Data were taken from (Grant 1967; Grant 1964; Neuman & Logan 1950; Fischer & Llaurado 1966; Harkness et al. 1957; Looker & Berry 1972; Farrar et al. 1965; Hosoda et al. 1984)

### 2. Quantitative parameters related to the fiber network arrangement

Along with a detailed assessment of the vascular biological composition, many studies have focused on the extraction of particular geometrical characteristics of the microstructure (Humphrey & Holzapfel 2012). Non exhaustively, the implemented image analysis methods focused on the analysis of the fiber diameters (D'Amore et al. 2010; Phillippi et al. 2014), the fiber lengths (D'Amore et al. 2010; Rezakhaniha et al. 2011; Hill et al. 2012; Tsamis, Phillippi, et al. 2013; Cicchi et al. 2014), fiber volume fractions (Tonar et al. 2003; Hayashi et al. 1974; O'Connell et al. 2008; Verheyen et al. 1987), as well as on the evaluation of fiber wavinesses (Hill et al. 2012; Roy et al. 2010; Rezakhaniha et al. 2012; Schrauwen et al. 2012), and orientations (Holzapfel 2006). More recently, for modelling purposes, more complex morphological features were quantified, such as fiber tortuosity (Koch et al. 2014), node connectivity and spatial intersections density (D'Amore et al. 2010; Koch et al. 2014), or density of transversely oriented segments (Koch et al. 2014). The number of elastic lamellae and of fenestrations inside the lamellae (Brüel & Oxlund 1996) and appearance of collagen fibers (Wolinsky & Glagov 1964) have also been investigated using image processing techniques.

# III. Characterization of the arterial tissue's mechanical function

Arteries exhibit a highly nonlinear mechanical behavior, which has long been investigated from different point of views. As a first approach to the arterial biomechanics, the macroscopic response of the arterial wall has been characterized through different uniaxial, biaxial, and tension-inflation tests along different physiologically-relevant directions. More recently, experimental setups coupling mechanical testing with live microscopy have been developed to decipher the microstructural mechanisms that are behind the non-linear character of the response. Finally, at the 100 nm scale, the constitutive responses of collagen and elastic fibers have been characterized. We here restrict the review to the *in vitro* mechanical testing techniques.

# A. Multiscale characterization techniques of the arterial structure

### 1. Macrostructure

Due to the biological nature of arterial tissue, the storage conditions and the temperature of test may have an impact on the in vitro mechanical response of the arterial tissue.

Concerning storage procedures, the arteries are usually harvested in freshly sacrificed animals or shortly after death. After excision, the arterial sample may be stored for 2-3 days in a saline solution at 4°C, or kept frozen at -20°C or at -80°C (Collins & Hu 1972; Pham et al. 2013), to avoid sample drying and accelerated sample degradation. The impact of these different protocols on the mechanical properties was investigated by comparing the uniaxial tensile response of specimen stemming from the same tissue sample, but subjected to different preparation protocols. As regards the temperature of test, two choices are classically made: either the ambient temperature of the room or the physiological temperature. Again, the uniaxial tensile responses relative to different temperature of tests are compared.

In order to characterize the arterial constitutive behavior, biomechanics imported the well-established uniaxial and biaxial tensile tests existing for metallic or inert materials to the arterial tissues: across the last seventy years, different arterial tissues stemming from different species and organs, at different ages, and different healthy or pathological states have been mechanically characterized. More precisely, for uniaxial tests, a dog bone sample is cut in the excised arterial segment, with the circumferential or longitudinal direction as main axis. Because uniaxial tensile tests do not reproduce the *in vivo* load conditions to which arteries are subjected, biaxial tests on flat squared samples have been developed. The load is applied simultaneously in both directions, maintaining a constant stress ratio between the two directions. Finally, tensile inflation tests make use of a cylindrical arterial segment, subjected to a fixed axial stretch and a variable internal pressure.

### 2. Microstructure

Several studies question the possible relation between the observed variability of the macroscopic mechanical response and the variability in the arterial wall composition or organization. By combining mechanical tests to histology or chemical access to the arterial composition, such studies seek for relations between arterial composition/morphology and subsequent mechanical response. The influence of each fiber network on the overall mechanical response has also been investigated: enzymes such as collagenase or elastase are used to (partially) degrade one constituent, and the mechanical response is then compared to the original one, and different degrees of degradation.

Besides, different studies proposed to physically separate the three arterial layers, so as to study the contribution of each arterial layer on the overall mechanical response (Weisbecker et al. 2012; Holzapfel et al. 2005; Sommer et al. 2010).

Aiming at deciphering the microstructural mechanisms that lead to the non-linear mechanical response, mechanical testing setups have been developed, permitting live imaging of the arterial microstructure with a confocal microscope during the application of the load. Such tests, coupled to the previously described image processing techniques, permit the characterization of the rearrangements of the different fiber networks.

### 3. Ultrastructure

At the ultrastructural scale, attempts were made to characterize the mechanical behavior of the elementary constituents of the arterial wall. We here focus on the collagen and elastin.

About collagen, the task is rendered more complex by the existence of different collagen types, with different mechanical properties. Our review is restricted to the characterization of collagen type I. This is the most abundant collagen type is bone, tendon, as well as in the tunica adventitia of arteries. Different characterization techniques have been employed for retrieving its mechanical characteristics at different length scales. Brillouin light scattering (Harley et al. 1977; Cusack & Miller

1979) gives access to the velocity of elastic waves having a wavelength of several hundreds of nanometers; subsequently, it allows characterizing the arrangement of several collagen molecules within fibrils. At a higher scale, nano-indentation tests performed by means of an Atomic Force Microscope provide access to the mechanical properties of collagen fibrils. Finally, the mechanical response of collagenous tissues at higher scales, from the fibrils to the tissue scales, is characterized by tensile tests, with different types and characteristic sizes of samples. The mechanical properties of collagen fibrils and fibers primarily depend on the hydration degree of the tissue under consideration.

The picture is quite different regarding the mechanical characterization of elastin or elastic fibers. Elastic fibers are often characterized by the mechanical properties of collagen-free biological tissues (such as arterial tissue or nuchal ligaments), considering that the ground substance surrounding the elastic fibers is not mechanically active. Collagen is removed either by autoclaving (Aaron & Gosline 1981) or by alkali treatment (Hass 1942b). Then, uniaxial tensile tests are performed on the extracted elastin fibers.

## B.   Universal hierarchical structure of the arterial tissue

The biological nature and the physiological functions of the arterial tissue render the characterization of the mechanical properties as a delicate task: harvesting the tissue implies the death of cells, with different consequences: in vitro characterization only investigates the passive response of the tissue, and therefore cannot account for the active role of smooth muscle cells in distributing the load across the tissue thickness; the stop in the constituent turnover and the progressive degradation of the organic constituents making up the tissue have consequences on the mechanical response of the tissue; and harvesting the tissue also implies the relaxation of some of the pre-stress and pre-stretch existing in the arterial tissue *in vivo*, leading to fiber rearrangements in the microstructure with consequences on the macroscopic mechanical response.

### 1.   Macrostructure

Before reviewing the macroscopic mechanical properties of arteries, attempts were made to characterize the influence of sample freezing or refrigerating on the resulting mechanical behavior. Several studies (Zemánek et al. 2009; Adham et al. 1996; Armentano et al. 2006) reported no significant variations of the mechanical response after storage, while, in other studies (Venkatasubramanian et al. 2006; Chow & Zhang 2011; Stemper et al. 2007), the variations in the mechanical behavior encompass variations in the initial and final stress-strain slopes, as well as changes in the knee point of stress strain curves, and in the ultimate stress. These variations may be explained by some damage occurring in the sample during freezing or refrigerating: formation of ice crystals, bulk water movement (Venkatasubramanian et al. 2006) can induce fiber cracking, loss of crosslinks, networks disruption, and death of cells. These variations may also be explained by the decrease in the collagen content after 48h cold storage (Chow & Zhang 2011), as well as by the exact procedure followed to freeze the sample. Still, it is impossible to decide on the directions of variations, since the different relevant studies come to apparently contradictory results. It is however generally admitted that freezing better maintains the mechanical properties of arterial samples than refrigeration (Chow & Zhang 2011; Stemper et al. 2007). (Zemánek et al. 2009) studied the influence of the temperature choice on the mechanical response of the arterial wall and showed that samples are stiffer at ambient temperature than at *in vivo* temperature, since a temperature increase by 1°C results in a 5% stiffness decrease; this result is in good agreement with (Fung 1993).

Furthermore, arteries are subjected *in vivo* to residual stresses and pre-stretch, as evidenced by (Bergel 1961; Fung 1993; Vaishnav & Vossoughi 1987), the amount of which depends on the organ and species under consideration; this *in vivo* stress-strain state originates in the growth and remodeling processes undergone by arteries and allows achieving a homeostatic stress state, being nearly uniform and equibiaxial across the arterial wall thickness (Humphrey 2009). As a consequence, excising and cutting open arterial segments partially release this existing *in vivo* stress-strain state, but there is no certainty that the load free configuration corresponds to a stress and strain free configuration.

Another important feature of the arterial constitutive behavior is the existence of a transient mechanical response: during the first mechanical cycles, the mechanical response of the arterial wall exhibits an important hysteresis which is reduced after several load cycles; the stabilized mechanical response barely shows any hysteresis. Experimentalists usually get rid of this transient response by performing several preconditioning cycles. The number of preconditioning cycles varies with the precise protocol, and the loading path and maximum load generally coincide with the further applied loading (see column 6 of Table 1). The microstructural underlying mechanisms occurring during preconditioning are not yet elucidated. The mechanisms are probably related to viscous effects, since the transient response is observed after a prolonged stop of the mechanical loading. Interestingly, (Zemánek et al. 2009) noticed that no preconditioning was necessary for equibiaxial tensile tests.

The previously described macroscopic testing procedures allow characterizing the mechanical response of a millimeter-sized arterial sample (see Table 1 for a literature review). The arterial wall exhibits a highly nonlinear mechanical response, which was already described in the 1880s (Roy 1881): while at low applied stresses arteries are very easily deformed, the arterial response becomes much stiffer at higher applied stresses. This nonlinear response occurs in any load direction. As many other biological tissues, arteries exhibit an important variability in their mechanical response, both across species, organs, location, or inter-individual (see Figure 5). As a result, there is an important variability in the stretch and stress levels at which the change of stiffness reaches its maximum.

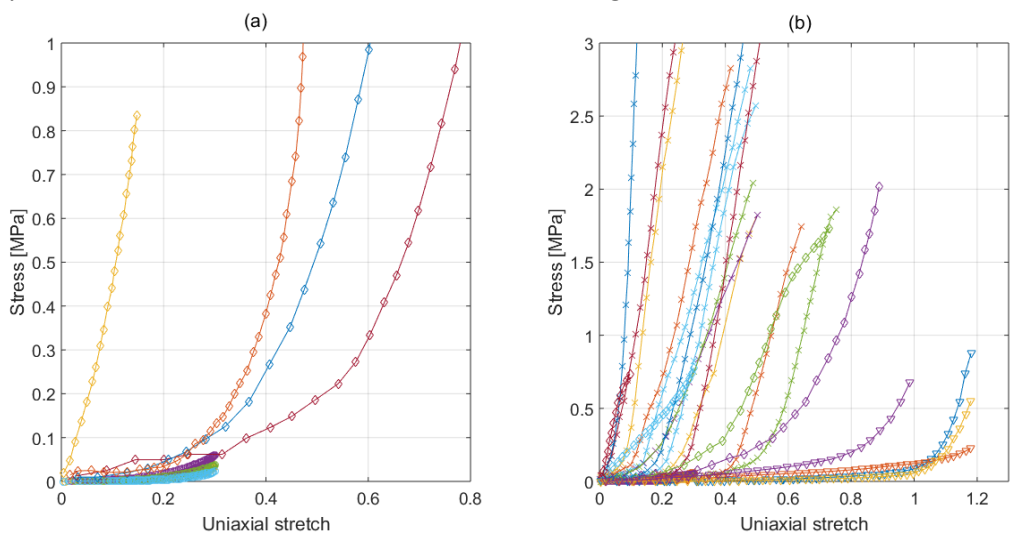

Figure 5: Mechanical response of different arterial tissues subjected to uniaxial tension in the (a) circumferential and (b) longitudinal direction: porcine coronary arteries (crosses, (Lally et al. 2004)), human ascending aorta (circles, (Choudhury et al. 2009)), human mid-thoracic descending aortas (diamonds, (Mohan & Melvin 1982)), and rabbit and pig aortas (triangles, (D. P. Sokolis et al. 2002; Sokolis et al. 2006)).

Concerning the tensile-inflation test, the arterial response is characterized by the variations of the arterial diameter as a function of the applied pressure as well as by the variations of the axial reaction force (Cox 1975; Weizsäcker et al. 1983; Dobrin 1986). The tensile-inflation tests evidence the salient feature of the *in vivo* pre stretch level; indeed, for axial pre stretches being smaller (resp. larger) than the *in vivo* pre stretch, the axial reaction force decreases (resp. increases) when the pressure increases. But, when the axial pre stretch is equal to the *in vivo* pre stretch, the axial reaction force does not depend on the applied inner pressure (Weizsäcker et al. 1983; Sommer et al. 2010) and remains constant during the pressure cycle (see Figure 6, $4^{th}$ curve from above). Similar results exist for isotonic tests, in which the axial force is kept constant but the axial pre stretch varies with the applied pressure (Cox 1975).

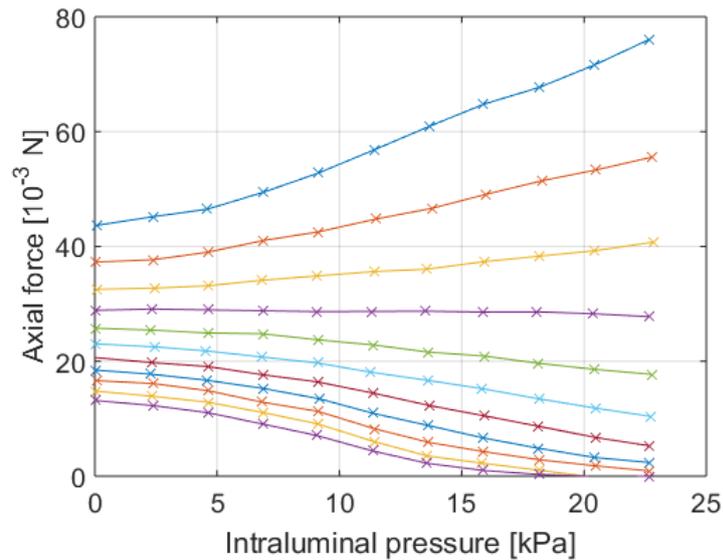

**Figure 6: Variation of the axial reaction force as a function of the applied luminal pressure, for different prescribed axial pre stretches. Experimental data from** (Weizsäcker et al. 1983)**.**

Anisotropy of arterial walls has also been investigated through these experimental setups (Vosshougi & Weizsäcker 1985; Lally et al. 2004). But no general conclusion can be drawn regarding the anisotropy of the arterial tissue. Still, carotid arteries are found stiffer in the circumferential direction (Cox 1975; Patel & Janicki 1970), while coronary arteries are stiffer in the longitudinal direction (Papageorgiou & Jones 1988; Patel & Janicki 1970). It is however remarkable that there is no statistical difference in the mechanical responses of arteries whether tested in the circumferential or in the longitudinal directions, when the tests are performed in conditions close to the physiological ones: this was observed e.g. by (Sato et al. 1979) on dog abdominal aortas, by (Dobrin 1986) on dog carotid arteries, and by (Sommer et al. 2010) on human carotid arteries.

Another much debated feature is the strain-rate dependence of the arterial mechanical response. It is now admitted that at low strain rates, the mechanical response of arteries does not vary with the strain rate (Sato et al. 1979; Zemánek et al. 2009; Tanaka & Fung 1974). The viscous character of arteries is however a much more complex question, since also creep and relaxation tests should be investigated.

Finally, these macroscopic mechanical tests also allow checking the widely accepted assumption of incompressibility of arterial tissues. Incompressibility implies that the changes in sample thickness can be deduced from the changes in the circumferential and axial dimensions. To our best knowledge, (Carew et al. 1968) was the only study that experimentally checked this assumption, by evaluating the ratio between bulk and shear moduli based on a tension-inflation test. As a result, the arterial bulk modulus was about three times larger than the Young's modulus determined by (Bergel 1961), while the hydrostatic and deviatoric stresses were of the same order of magnitude. They could therefore conclude that arteries are only slightly compressible.

| Reference | Animal | Artery | Layer | Sollicitation | Preconditioning | Direction | Number of tests |
|---|---|---|---|---|---|---|---|
| (Balzani et al. 2006) | Human | Abdominal aorta | Media | Uniaxial tensile test on strips in saline solution at 37°C | 5 loading-unloading cycles at 1 mm.min$^{-1}$ | LONG | 1 |
| (Balzani et al. 2006) | Human | Abdominal aorta | Media | Uniaxial tensile test on strips in saline solution at 37°C | 5 loading-unloading cycles at 1 mm.min$^{-1}$ | CIRC | 1 |
| (Choudhury et al. 2009) | Human | Ascending thoracic aorta | All | Equibiaxial test on flat square samples | | | 5 healthy * 4 locations |
| (Chuong & Fung 1984) | Rabbit | Thoracic aorta | All | Compression on square flat samples in air | Loading until 600g and then unloading | RADIAL | 4 |
| (Cox 1975) | Dog | Carotid arteries | All | Inflation tests at constant tensile stretch or constant tensile load, controlled temperature | Various number of inflation/deflation cycles until a stabilized response is found | CIRC at fixed LONG stretch And CIRC at fixed LONG force | 10 animals |
| (Dobrin 1986) | Dog | Carotid arteries | All | Inflation test at fixed tensile stretch | 5 inflation/deflation tests until stabilization of the response | CIRC at fixed LONG stretch | 120 tests |
| (Hill et al. 2012) | Rabbit | Carotid arteries | All | Tension of flat dog bone samples in a saline solution | 5 cycles at 1N under quasi static loading | CIRC | 8 samples |
| (Holzapfel 2006) | Human (80 years old woman) | Abdominal aorta | Adventitia/media/intima | Tensile test in a saline solution on strips | 5 cycles preconditioning | CIRC and LONG | 2 per layer |
| (Holzapfel et al. 2005) | Human | Left anterior coronary artery | Adventitia / media / intima | Tensile test in a saline solution on strips | 5 cycles at 1mm/min | CIRC and LONG | 78 strips (6 * 13) |
| (Keyes et al. 2013) | Porcine | Descending coronary arteries | All | Planar biaxial testing at different stress ratios and tubular testing | 10 cycles tensile inflation tests | LONG/CIRC | 5 |
| (Kim & Baek 2011) | Porcine | Thoracic Aorta | All | Inflation tests under fixed axial stretch | In the longitudinal and circumferential direction | CIRC at LONG | 7 |
| (Lally et al. 2004) | Porcine | Coronary arteries | All | Tensile tests on strips in a saline bath | 5 cycles at 1N at a strain rate of 60%/min | LONG | 21 |
| (Lally et al. 2004) | Porcine | Coronary arteries | All | Equibiaxial tests on square | 5 cycles at 0.5N at a strain | LONG/CIRC | 8 |

| Reference | Species | Location | Layer | Test | Preconditioning | Direction | N |
|---|---|---|---|---|---|---|---|
| | | | | samples in a saline bath | rate of 60%/min | | |
| (Mohan & Melvin 1982) | Human | Mid-thoracic descending aorta | All | Tensile tests at room temperature sprayed with saline solution | 5 cycles | LONG and CIRC | 18 each |
| (Patel & Janicki 1970) | Dog | Coronary and carotid artery | All | Inflation test at *in vivo* stretch | Response stabilizes after two load cycles | LONG/CIRC | 14 animals |
| (Pandit et al. 2005) | Porcine | Coronary arteries | All | Inflation test in saline solution at imposed axial stretches | Not reported | CIRC | 12 |
| (Papageorgiou & Jones 1988) | Human | Iliac arteries | All | Inflation test at *in vivo* length, static tests, kept wet; followed by axial tensile test on the whole sample | One cycle pressure up to 160 kPa | CIRC followed by LONG | 18 |
| (Samila & Carter 1981) | Human | Carotid arteries | Adventitia removed? | Uniaxial tensile tests in saline solution | Not reported | LONG and CIRC | 20 |
| (Sato et al. 1979) | Dog | Abdominal aorta, carotid arteries, femoral arteries | All | Tensile test at fixed pressure, 37°C, sample kept wet | Repeated pressure loading and cyclic stretching | LONG at fixed CIRC | 5 animals |
| (Schmid et al. 2005) | Human | aorta | adventitia | Tensile test immerged in a saline bath | 3 quasi static loading /unloading cycles | LONG and CIRC | 2 samples |
| (Silver et al. 2003) | Porcine | Aorta, carotid and vena cava | All | Tensile test in a saline bath on strips | Not reported | LONG and CIRC | 6 each |
| (Dimitrios P. Sokolis et al. 2002) | Rabbit | Abdominal aorta | media | Tensile test on strips at 37°C immerged in a saline solution | 10 tensile cycles to the maximum tensile strain | LONG | 15 |
| (Dimitrios P. Sokolis et al. 2002) | Porcine | Abdominal aorta | media | Tensile test on strips at 37°C immerged in a saline solution | 10 tensile cycles to the maximum tensile strain | LONG | 20 |
| (Sokolis et al. 2006) | Rabbit | Descending thoracic aorta | all | Tensile test on strips at 37°C immerged in a saline solution | 10 tensile cycles to the maximum tensile strain | LONG | 35 |
| (Sommer et al. 2010) | Human | Carotid arteries | All and layer specific | Inflation under controlled axial force | 5 cycles axial preconditioning followed by 5 inflation-deflation test | CIRC | 20 |

| Reference | Species | Location | Layer | Test | Loading | Direction | Samples |
|---|---|---|---|---|---|---|---|
| (Storkholm et al. 1997) | Porcine | Abdominal aorta | All | Inflation test under controlled axial stretch | Cyclic loading/unloading test between 0 and 25 kPa until stable response | CIRC | 5 |
| (Tanaka & Fung 1974) | Dog | Aorta | All | Uniaxial tensile test on strips in a saline solution | Cyclic loading unloading until stabilization | LONG and CIRC | |
| (Vande Geest et al. 2006) | Human | Abdominal aorta | All | Flat biaxial tests at different stress ratios | 9 loading/unloading cycles for each tension ratio | LONG/CIRC | 8 healthy specimens |
| (von Maltzahn et al. 1984) | Bovine | Carotid arteries | Intact and media/intima only | Inflation tests under different stretch ratios | Inflation/deflation tests between 0 and 250 mmHg at 1 kPa.s$^{-1}$ | CIRC | 18 |
| (Vorp et al. 2003) | Human | Ascending aorta | All | Uniaxial tensile in a saline solution | Not reported | LONG and CIRC | 7 each |
| (Vosshougi & Weizsäcker 1985) | Rat and pig | Aorta | Media and Intima | Uniaxial tensile test in a saline solution at 37 and 24°C | Three loading/unloading cycles at constant extension rate | LONG and CIRC | 13 rats and 7 pigs |
| (Weisbecker et al. 2013) | Human | Thoracic aorta | Media | Uniaxial tensile test in a saline solution at 37°C | Not reported | CIRC | 1 sample not elastase treated |
| (Weizsäcker et al. 1983) | Rat | Carotid arteries | All | Inflation test at constant axial stretch at 37°C, in a saline bath and axial stretch at different pressures | 10 inflation/deflation tests | LONG and CIRC | 6 rats |
| (Zeinali-Davarani et al. 2015) | Porcine | Descending thoracic aorta | All | Planar biaxial tests at different stress ratios | Eight loading cycles up to 30 N.m$^{-1}$ | LONG/CIRC | 4 locations, 1 sample each |
| (Zemánek et al. 2009) | Porcine | Thoracic aorta | all | Equibiaxial tensile test in a saline solution | Cyclic loading of the specimen at 0.5 N until stabilization of the response | LONG/CIRC | 4 locations, 1 sample each |

Table III.1: Uniaxial, biaxial, and tension inflation tests on large elastic arteries. LONG (resp. CIRC) stands for the longitudinal (resp. circumferential) direction

## 2. Microstructure

As a transition between purely mechanical characterization and mechanical testing coupled with live imaging, we here focus on correlations existing between mechanical characteristics of arteries and composition data. In this respect, we report two investigation results.

Firstly, several studies (Sato et al. 1979; Hayashi et al. 1974) laid emphasis on the dependence of the mechanical response on the sample location along the aortic tree. The aortic stiffness is found to be higher in distal regions as compared to proximal regions (Zeinali-Davarani et al. 2015; Sokolis 2007; Haskett et al. 2010; Tanaka & Fung 1974), which correlates with a higher collagen content in the distal aortic regions. This property is directly related to the difference in mechanical function between proximal and distal regions: proximal regions of the aortic tree directly receives blood from the heart and therefore needs to exhibit larger damping properties, which is microstructurally translated through a larger elastin content and more undulated collagen bundles in the proximal aortas (Zeinali-Davarani et al. 2015).

Secondly, several mechanical tests were performed after partial enzyme degradation of either elastin or collagen. By imparting compressive stresses to the collagen (Chow et al. 2014), the presence of elastin increases the collagen folding, resulting in a more compliant response of the tissue (Ferruzzi et al. 2011), since straightened collagen fibers are stiffer than elastin fibers. As a consequence, degrading elastin leads to vessel enlargement, by relaxation of the internal compressive stress. This results in a mechanical response being softer in the low stress regime, and stiffer in the large stress regime (Fonck et al. 2007; Weisbecker et al. 2013). Elastin degradation also leads to an earlier recruitment of collagen fibers, the mechanical response being sooner stiffer (Rezakhaniha et al. 2011; Zeinali-Davarani et al. 2013). On the other hand, collagen degradation leads to the disappearance of the progressive stiffening of the mechanical response, while the initial slope of the stress-strain curves remains unchanged (Weisbecker et al. 2013).

Besides, arteries exhibit a layer-specific mechanical response, which depends on the layer morphology. Mechanical tests on the tunica intima were performed by (Holzapfel et al. 2005; Weisbecker et al. 2012): intima exhibits a stiffer mechanical response when loaded in the longitudinal direction than in the circumferential one, which is in good agreement with the longitudinal orientation of the fibers of the internal elastic lamina (Farand et al. 2007). However, it is generally agreed on, that the tunica intima barely contributes to the mechanical response of arteries. As regards the tunica media, the circumferential direction shows a stiffer uniaxial response than the longitudinal direction, which correlates with the preferred circumferential orientation of the fiber networks in the media. Contrarily, in the tunica adventitia, the uniaxial mechanical response is stiffer in the longitudinal direction than in the circumferential one (Holzapfel et al. 2005; Weisbecker et al. 2012), because of the longitudinal orientation of the fibers of the strips in the load-free configuration (Chen et al. 2013), see Figure 7. Finally, the larger elastin content in the media as compared to the adventitia makes the media more compliant, while the adventitia resists to larger loads.

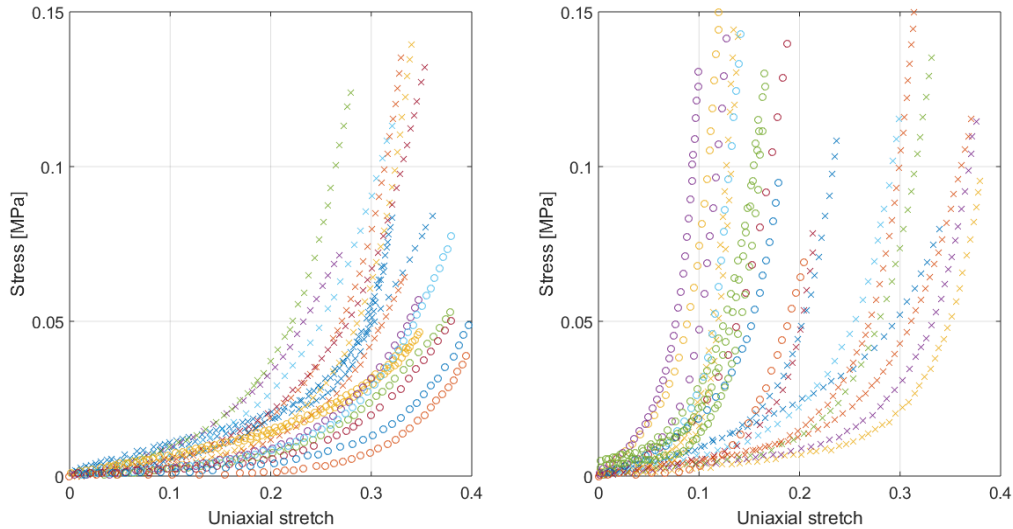

Figure 7: Mechanical response of the tunica media (a) and adventitia (b) for uniaxial tests performed in the circumferential direction (crosses) and in the longitudinal direction (circles). The engineering stress is plotted against the uniaxial stretch, for human descending coronary arteries (Holzapfel et al. 2005), human and porcine aortas (Balzani et al. 2006; Schmid et al. 2005).

At the scale of the fiber networks, the microstructural origin of the nonlinear mechanical behavior could be evidenced by means of confocal microscopy: namely, in the course of the mechanical loading, the fiber networks making up the microstructure tend to rearrange to better resist the loading. In situ mechanical testing showed the ability of the different fiber networks to resist the applied mechanical loading by progressively aligning with the load direction (see Figure 8). At rest, the collagen bundles of the tunica adventitia form a dense and crimped network, which is progressively straightened and then stretched in the course of the applied loading. The straightening mechanism of fibers is generally referred to as the fiber recruitment and coincides with the response at low stresses (Hill et al. 2012; Schrauwen et al. 2012; Sokolis et al. 2006): as long as collagen fibers are not straight, they cannot sustain load and the softer constituents making up the arterial tissue sustain the applied load. Furthermore, whatever the load direction in both uniaxial and biaxial tests, adventitial collagen bundles also have the faculty to reorient and to align with the load direction (Chen et al. 2013); the collagen bundles stretching takes place after the fiber realignment, as shown by the tracking of the collagen bundle deformation, either by means of fluorescent microspheres (Chen et al. 2011), or by means of X-ray diffraction (Schmid et al. 2005). Nevertheless, the mechanisms driving the collagen network reorientation remains unclear. Two hypotheses are currently discussed: the affine rotation, where the fibers follow the matrix deformation (Wan et al. 2012); or larger rotations of the collagen network, where the collagen bundles are able to generate shear stresses to rotate faster than the matrix (Jayyosi 2015; Billiar & Sacks 1997). It might well be that both assumptions are valid but in different strain ranges, the affine rotations being validated over the physiological deformation range (Wan et al. 2012). The collagen bundles recruitment is driven by the presence of the elastin network: in the adventitia, the elastin network is at rest aligned with the collagen network (Chen et al. 2011); and it tends to reorient and align by application of a mechanical loading. Still, the realignment of the adventitial elastin network is less pronounced that the collagen realignment (Chen et al. 2013).

In the media, elastin lamellae, collagen fibers, and smooth muscle cells also undergo load-induced reorientation. Under uniaxial and biaxial load cases, the medial fiber networks and the collagen network tend to align with the load direction (Timmins et al. 2010). Also, the engagement of collagen fibers occurs first in the media, and starts later in the adventitia (Zeinali-Davarani et al. 2015; Chow et al. 2014). However, the amplitude of fiber rotation is the largest for the collagen bundles of the adventitia (see Figure 8): rotations of the medial constituents as well as of the adventitial elastin remains limited (Chow et al. 2014); this may be related to the very different nature of the networks, the elastin network of the media being for instance a very dense and interconnected network, with elastic segments oriented in all directions. At a larger scale, the elastic lamellae progressively unfold with the load application and then stretch, as observed by means of polarized light microscopy by (Sokolis et al. 2006). The cohesive pericellular interlaced bundles also straighten and reorient by application of a load (Sokolis et al. 2006), and this recruitment process was shown to be faster than the recruitment of the circumferentially oriented parallel bundles covering the elastic lamellae (Sugita & Matsumoto 2016); the latter authors propose the stiffness difference between elastin and smooth muscle cells as the possible explanation, the more compliant surrounding medium allowing faster recruitment of the initially crimped fibers.

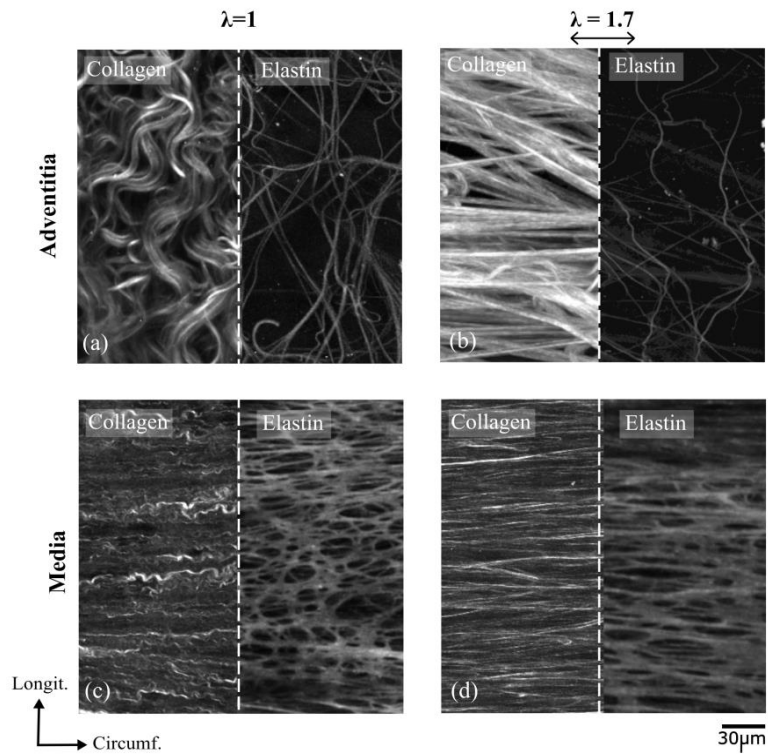

Figure 8: Load-induced changes in the morphology of the fiber networks under a circumferential uniaxial load case. Top: adventitial networks; Bottom: medial networks; Left: load-free configuration; Right: uniaxial stretch of 1.7. Images taken on a rabbit carotid artery with a confocal bi-photon microscope (IVTV Platform, ANR-10-EQPX-06-01, FR).

### 3. Ultrastructure

In this section, the mechanical properties of the collagen and elastin structures are reviewed.

Concerning collagen, the literature reports a very broad range of variations for the mechanical properties of collagenous tissues: from several hundreds of MPa for a hydrated arrangement of collagen fibers to a few tens of GPa for an arrangement of air-dried collagen molecules (see Table 2 for a literature review). At an intermediate scale, collagen fibrils have an elastic modulus of several GPa, whereby this exact value depends on the hydration degree of the sample.

| Reference | Origin of the tissue | Lengthscale | Method | Hydration state | Measured stiffness |
|---|---|---|---|---|---|
| (Harley et al. 1977) | Rat tail tendon | Molecular | Brillouin light scattering | Variable hydration states | 9-21,5 GPa |
| (Cusack & Miller 1979) | Rat tail tendon | Molecular | Brillouin light scattering | Variable hydration states | 7,8-17,9 GPa |
| (Eppell et al. 2006) | Dermis of sea | Molecular (nanofibril) | Tensile test using a microdevice | Hydrated | 6 GPa |
| | cucumber | | | | |
| (Lorenzo & Caffarena 2005) | Molecular dynamics | Molecular | Molecular dynamics | | 4,8 ± 1 GPa |
| (Strasser et al. 2007) | Calfskin | Fibrils | Nano indentation by AFM | Dried | 1.2 GPa (transverse direction) |
| (Yadavalli et al. 2010) | Calfskin | Fibrils | Nano indentation by AFM | Dried | 1.03 ± 0.31 GPa (tranverse direction) |
| (van der Rijt et al. 2006) | Bovine Achilles tendon | Fibrils | Tensile test using AFM | Dried | 2-7 GPa (Young's modulus) |
| | | | | Hydrated | 0,2-0,8 GPa (Young's modulus) |
| (Kato et al. 1989) | Rat tail tendon | Fibrils | Tensile test | Air-dried overnight | 2.200 ±0.5 GPa |
| | | | | Wet state | 500 ± 150 MPa |
| (Sasaki & Singo Odajima 1996) | Bovine Achilles tendon | Fibrils | X-ray diffractometry | In a saline solution | 2,9 GPa (Young's modulus) |
| (Shen et al. 2008) | Dermis of sea cucumber | Fibrils | Uniaxial tension using a microelectromechanical system | Hydrated | 0,86 ±0,45 GPa (Young's modulus) |
| (Svendsen & Thomson 1984) | Rat tail tendon | Fibrils | Tensile test | Hydrated | 1.2-1.8 GPa |
| (Wenger et al. 2007) | Rat tail tendon | Fibrils | Nano indentation by AFM | Different hydration states | 3.75-11.5 GPa |
| (Yang et al. 2008) | Bovine Achilles tendon | Fibrils | Bending experiment by AFM | Hydrated | 1.0 – 3.9 GPa |
| (Sasaki & Shingo Odajima 1996) | Bovine Achilles tendon | Fibrils | Tensile test | Hydrated | 445 MPa |
| (Silver et al. 2001) | Turkey tendons | Tendon | Tensile test | Hydrated | 62 MPa |

Table III.2: Mechanical properties of collagen: from molecular arrangement to collagenous tissues.

Concerning elastin, its mechanical behavior is purely elastic, with no hysteresis nor transient effect up to more than 100% strain, eventually showing a brittle failure mode (Aaron & Gosline 1981). The Young's moduli reported in the literature exhibit less variability: the elastic modulus of intact purified elastic fibers from aortas was reported to be 0.4 MPa (Hass 1942b; Hass 1942a; Hass 1943; Krafka 1937; Burton 1954; Faury 2001), while the water-swollen, single elastin fibers of (Aaron & Gosline 1981) has a Young's modulus of 1.2 MPa. According to (Gundiah et al. 2007), the difference in the Young's modulus could be explained by the method for elastin extraction, the autoclaving method, as used in (Aaron & Gosline 1981), providing higher stiffness values as compared to alkali-treated elastin, as proposed by (Hass 1942b).

Finally, smooth muscle cells do not contribute at rest to the mechanical properties of arteries, with an elastic modulus of a few tens of kPa and a high (up to 150%) distensibility (Faury 2001; Nagayama & Matsumoto 2004). Their functional role is however of prime importance, since they are responsible for the regulation and maintenance of blood flow, to permit a more constant peripheral blood flow (Ratz 2004; Faury 2001). There is also a broad consensus that smooth muscle cells distribute the load across the arterial thickness: single lamellar units are able to bear a certain tension of 1-3 $N.m^{-1}$ (as calculated by the product of the intraluminal pressure by the vessel radius), independently of organs or species under consideration (Faury 2001; Humphrey 2008; H Wolinsky & Glagov 1967). At the nanoscale, several proteins are considered as influential to vascular smooth muscle cells mechanics, namely: (i) the motor protein, called myosin II, assembled into thick filaments; (ii) filamentous actin, which can be seen as a cable on which myosin heads bind and can "walk" on (Ratz 2016); and (iii) smooth muscle titin and microtubules of the cytoskeleton. Actin filaments weave through the smooth muscle cell and are part of its contractile apparatus. They consist in semi-flexible polymers that can, in conjunction with myosin, act as biological active springs able to exert or resist against force in a cellular environment (Blanchoin et al. 2014). They therefore participate in arterial acute adaptive plasticity (Bednarek et al. 2011). It has also been shown that smooth muscle cell filamentous actin is in a continuous state of remodeling (Bursac et al. 2007). Other studies about myosin thick filaments of muscle cells have revealed their partial dissociation and reformation during, respectively, relaxation and contraction (Smolensky et al. 2005).

# IV. Conclusion

This article demonstrates the great effort which has been put to reveal the hierarchical character of arterial tissues, from both a structural and a mechanical point of view. Healthy arterial tissue appears as a very complex tissue, with a local composition and structure perfectly suited to the mechanical and physiological function of the tissue at this precise location: cells are able to maintain homeostatic mechanical properties by adjusting the local concentrations of the diverse constituents. Interestingly, the present study demonstrates the existence of a strong correlation between elastin and collagen content in healthy arterial tissue. Still, the complexity in the arterial structure and composition leads to an important variability of the mechanical response of arteries, as sketched in the different tables and figures of this article. The question arises to which extent this variability may be explained by the local variations in the composition and constituents' arrangement. To answer this question, two strategies need to be developed in parallel, namely a vast experimental campaign and the development of an effectively multiscale constitutive model for arteries. From an experimental point of view, there is a pressing need for experimental data coupling composition data with microscopic observations and mechanical testing for the same arterial segment; such data is necessary for a better understanding of the different mechanisms occurring within the arterial wall, but also for validating the multiscale constitutive model. These mechanisms may also be better deciphered when focusing on the evolution of the arterial mechanical properties due to remodeling. From the theoretical point of view, there already exist numerous constitutive models for arteries, with varying complexity. These models are very well-suited to powerful computational approaches of the arterial biomechanics. However, these models remain phenomenological in nature and therefore are not meant to account for the microscopic mechanisms occurring in arteries. As a consequence, any (compositional or structural) change in the microstructure can only be accounted for by a new determination of the model parameters. The proposed multiscale approach involves a deep change of perspective: the arterial tissue is understood as a heterogeneous material; its mechanical properties can then be estimated from the mechanical behavior of the inhomogeneities, considered as homogeneous phases, their fractions, their characteristic shapes, and their interactions within a well-chosen volume called the representative volume element. The mechanical properties of each phase can themselves arise from a homogenization process over a representative volume element containing heterogeneities with smaller characteristic dimensions. Such an approach would allow to effectively relate the arterial composition and structure to the mechanical behavior, as reviewed in this article; and therefore to discriminate between variability and effect of the composition and structural arrangements. This multiscale approach could be also especially useful when focusing on damage and strength of arterial wall, which is of prime importance for medical applications.